\documentclass[12pt]{iopart}
\usepackage{epsfig}
\def\Journal#1#2#3#4{{#1} {\bf #2} (#4) #3}
\def\AP{\em Ann. Phys.}

\def\NPA{{\em Nucl. Phys.} A}

\def\PLB{{\em Phys. Lett.} B}

\def\PRL{\em Phys. Rev. Lett.}

\def\ZP{\em Z. Phys.}
\newcommand{\fel}{\frac{1}{2}}
\newcommand{\harm}{\frac{1}{3}}
\newcommand{\negy}{\frac{1}{4}}
\newcommand{\dmu}{\partial^{\mu}\!}
\newcommand{\dmuu}{\partial_{\mu}\!}
\newcommand{\dnu}{\partial^{\nu}\!}
\newcommand{\dnuu}{\partial_{\nu}\!}
\newcommand{\msun}{\, M_{\odot}}
\newcommand{\okgr}{\Omega_{\rm K}}
\newcommand{\nuk}{\nu_{\rm K}}

\begin{document}
\title[Neutron star properties]{Neutron star properties with in-medium vector 
mesons}
\author{F. Weber\P, Gy.~Wolf\dag\footnote[2]{Present address: KFKI, 
RMKI, H-1525 Budapest, POB.~49, Hungary}, 
T. Maruyama\dag\footnote[4]{College of Bioresource Sciences, Nihon University, 
Fujisawa, Kanagawa-ken 252-8510, Japan\\
Atomic Energy Research Institute, College of Science and Technology,
Nihon University, 1-8-14 Surugadai, Kanda, Chiyoda-ku 101-8308 Tokyo,
Japan} and S. Chiba\dag}
\address{\dag\ Japan Atomic Energy Research Institute, \\
Tokai, Naka, Ibaraki 319-1195, Japan}
\address{\P University of Notre Dame, Department of Physics, 225
Nieuwland Science Hall Notre Dame, IN 46556-5670, USA\\ 
\bigskip
Email:~ {\tt wolf@rmki.kfki.hu}, {\tt fweber@nd.edu}}
% \end{small}

\begin{abstract}
We explore the impact of in-medium modification of the properties of
vector mesons on the nuclear equation of state and neutron star
properties. It is found that in-medium modifications stiffen the
nuclear equation of state considerably. If this feature has its
correspondence in the full treatment of dense hadronic matter, then
very little room is left for the existence of exotic phases like quark
matter or boson condensates in the centers of neutron stars of
canonical mass.
\end{abstract}

Submitted to: {\it J.\ Phys.\ G: Nucl.\ Phys.}


\maketitle

\section{Introduction}\label{sec:intro}

The exploration of the properties of in-medium hadrons is a topic of
very great current interest. This is reflected by the number of
experiments either planned or recently performed which target at the
determination of medium modification of hadrons, especially of $K^\pm$
and $\rho$ mesons. The first results of the CERES and HELIOS
collaborations show a significant enhancement of the production of
low-mass lepton pairs \cite{CERES,Helios3} in relativistic
nucleus-nucleus collisions, compared to the yield that one expects
from proton-proton collisions.  Since these lepton pairs are direct
signals from the decay of vector mesons within the hot reaction zone,
this enhancement points toward modifications of the masses and/or
widths of vector ($\rho$ and $\omega$) mesons in a dense hadronic
environment \cite{likobrown,Brown-Rho,RCW}.

Medium modifications of mesons, as described just above, are generally
ignored in standard treatments of hot and dense hadronic matter
\cite{weber99:book}. In the framework of such treatments, the nuclear
interaction is described through the exchange of mesons whose
properties are entirely medium independent.  The most important of
these mesons are the $\pi$, $\sigma$, $\rho$ and $\omega$ mesons. It
seems to be a general feature that standard treatments fail to
reproduce the observed enhancement in the production of low-mass lepton
pairs~\cite{Drees}. This appears to be different for theoretical
treatments which account for in-medium modifications of vector mesons.

In this paper we introduce two such treatments, one based on the
Brown-Rho scaling scheme while the other constitutes a novel approach
based on the techniques of effective nuclear field theory, which are
used to computed corrections to the nuclear equation of state (EOS)
that originate from the in-medium modifications of $\rho$ and $\omega$
vector meson properties. These models for the equation of state are
then used to determine several key properties of neutron stars, such a
the limiting stellar mass, the mass-radius relationship, and the
Keplerian velocity. These quantities are known to be very sensitive to
variations in the nuclear equation of state \cite{weber99:book} and,
therefore, serve to test as to whether the in-medium vector-meson
modifications computed for the two theoretical treatments are
compatible with solid astrophysical data on neutron stars.

\section{Theoretical treatment of vector mesons in matter}

\subsection{Choice of Lagrangian}

In order to study the medium modifications on vector mesons, described
in section \ref{sec:intro}, we assume that the interactions producing
these modifications are independent of the interactions used in the
Walecka lagrangian, so we may take them into account by only modifying
the meson properties themselves.  For the nuclear interaction we adopt
the non-linear Walecka model. In this model the basic degrees of
freedom are the nucleon $N$, the scalar self-interacting $\sigma$
meson, and the $\omega$ and  $\rho$ vector mesons. The lagrangian
is given by
\begin{eqnarray}
 {\cal L} &=&\bar\Psi \gamma^\mu (i\partial_\mu - g_\omega \omega_\mu)\Psi
              - \bar\Psi (m_N-g_\sigma \sigma) \Psi \nonumber \\
            && -\negy (\dnu \,\omega^\mu - \dmu \, \omega^\nu)
               (\dnuu \, \omega_\mu - \dmuu \, \omega_\nu)
              +\fel m_\omega^2 \omega_\mu \omega^\mu \nonumber \\
            && + \fel (\dnuu \, \sigma \dnu \, \sigma  - m_\sigma^2 \sigma^2)
              - \harm b m_N (g_\sigma \sigma)^3 -
              \negy c (g_\sigma \sigma)^4 \nonumber \\
            && -\negy (\dnu {\bf \rho}^\mu - \dmu {\bf \rho}^\nu)
               (\dnuu \, {\bf \rho}_\mu - \dmuu \, {\bf \rho}_\nu)
              +\fel m_\rho^2 {\bf \rho}_\mu {\bf \rho}^\mu \nonumber \\
            && -g_\rho {\bf \rho}_\nu (\fel \bar\Psi \gamma^\nu {\bf
\tau}\Psi
              + {\bf \rho}_\mu \times (\dnu {\bf \rho}^\mu-\dmu {\bf
\rho}^\nu)
              + 2 g_\rho ({\bf \rho}^\nu \times {\bf \rho}^\mu)
                  \times {\bf \rho}_\mu) \, ,
\label{eq:lag}
\end{eqnarray}
where $\Psi$ denotes the nucleon fields, and $g_\sigma$, $g_\omega$,
$g_\rho$, $b$, and $c$ are the meson-nucleon coupling constants
\cite{weber99:book}. The equations of motion that follow from
(\ref{eq:lag}) are solved in the mean-field approximation, where the
meson fields are substituted by their classical expectation
values. The parameters of the model ($g_\sigma$, $g_\omega$, $g_\rho$,
$b$, and $c$) are fitted to the binding energy ($-16$~MeV), nuclear
incompressibility ($K=240$~MeV) and effective nucleon mass ($m^*_N =
0.8 \; m_N$) at the saturation density ($\rho_0 = 0.168~{\rm
fm}^{-3}$) of infinite nuclear matter.  (A soft equation of state was 
predicted recently by a novel many-body approach that is based on the 
principles of chiral perturbation theory \cite{Lutz}.)
In passing we mention that we
ignore possible medium modifications of the properties of the scalar
$\sigma$ meson. Being a Goldstone boson, it is not clear how the
properties of this meson may change in dense hadronic
matter. Application of the Brown-Rho scaling scheme to the
$\sigma$-meson mass, for instance, leads to an over-binding of nuclear
matter \cite{Rapp}. Since we are mainly interested in the high-density
part of the equation of state of hadronic matter, which is dominated
by the exchange of vector mesons among the nucleons, it is rather save
for our purposes to ignore medium effects on the $\sigma$ meson.

\subsection{In-medium modifications of vector meson properties}
\label{sec:schemes}

Next we turn to the two alternative models employed in this paper to
explore the in-medium properties of $\rho$ and $\omega$ mesons in
dense hadronic matter.  The first model is based on the familiar
Brown-Rho scaling \cite{Brown-Rho} according to which the masses of
vector mesons and nucleons in a hadronic medium scale as
\begin{equation}
\frac{m_\omega^*(\rho)}{m_\omega} =\frac{m_\rho^*(\rho)}{m_\rho} =
\frac{m_N^*(\rho)}{m_N} \, ,
\end{equation}
where the asterisks denotes the particle masses in the medium.  For
the motivation of this scaling law we refer to the original literature
\cite{Brown-Rho}.  The effective mass of the nucleon at a given
density, $m_N^*(\rho)$, is taken from the Walecka model,
\begin{equation}
  m_N^*(\rho) = m_N - g_\sigma <\sigma> \, ,
\end{equation}
with $<\sigma>$ the mean-field ground-state expectation value of the
$\sigma$ meson.  The second model which we study here \cite{LWF} is a
novel approach based on the techniques of effective nuclear field
theory.  In the framework of this model, a low-density expansion is
performed in order to establish a connection between the meson-nucleon
scattering amplitudes and the properties of mesons in nuclear
matter~\cite{LDT}. This scheme leads to a systematic expansion for the
in-medium self energy of a meson (as well as any other particle) in
terms of the corresponding vacuum scattering amplitudes. The
vector-meson--nucleon scattering amplitudes are not directly related
to experimental observables. Instead they are determined indirectly in
a coupled-channel scheme where the $\rho N$ and $\omega N$ channels
enter in intermediate and final states of measured processes. The
parameters of this theory are fixed by fitting the available
meson-nucleon scattering data in the relevant energy range. We employ
these scattering amplitudes to construct the vector-meson self
energies in nuclear matter to leading order in density.

The effective field theory provides us with the scattering lengths and
spectral functions of vector mesons as a function of the density. For
$\rho$ and $\omega$ vector mesons, the $\rho$--$N$ and $\omega$--$N$
scattering lengths are defined as ($V = \rho, \omega$)
\begin{equation}
 a_{VN}=f_{VN}(\sqrt{s}=m_N + m_V) \, ,
\end{equation}
which leads to
\begin{equation}
a_{\rho N} =(-0.1+0.6\, {\rm i}) ~{\rm fm} ~~ {\rm and}  ~~ 
a_{\omega N}= (-0.5+0.2\,{\rm i}) ~{\rm fm} \, . 
\end{equation}
To lowest order in density, this corresponds to the following
in-medium modifications of the vector meson masses and their widths at
nuclear matter density:
\begin{equation}
 \Delta m_\rho \simeq 10~{\rm MeV}\, ,~~ \Delta m_\omega \simeq
50~{\rm MeV} \, , ~~ \Delta\Gamma_\rho \simeq 120~ {\rm MeV} \, , ~~
\Delta \Gamma_\omega \simeq 40~{\rm MeV} \, .
\end{equation}
We stress that the coupling of the vector mesons to baryon resonances
below threshold, which is reflected in the strong energy dependence of
the scattering amplitudes, cannot be neglected. One should therefore
use the whole vector-meson spectral functions as done in this paper in
order to achieve a most reliable description. As it turns out, however,
application of the simple mass shift scheme for the in-medium effects
on vector mesons leads to results that are rather similar to those
obtained by using the full treatment, based on medium dependent
spectral functions.

As a side-remark we mention that the validity of these two schemes
adopted to include in-medium modifications in the equation of state is
naturally limited to moderately compressed hadronic matter. The extent
to which they are applicable to very highly compressed hadronic matter
as existing in the centers of very heavy neutron stars is an open
issue.  The situation looks quite promising, though, for neutron stars
of average mass, $M \simeq 1.4 \, \msun$, whose central densities may
be just a few times higher than the density of ordinary nuclear
matter, depending of the stiffness of the equation of state (cf.\
section \ref{sec:nss}). 

\section{Numerical outcome}

\begin{figure}[tb]
\begin{center}
\leavevmode
\epsfig{file=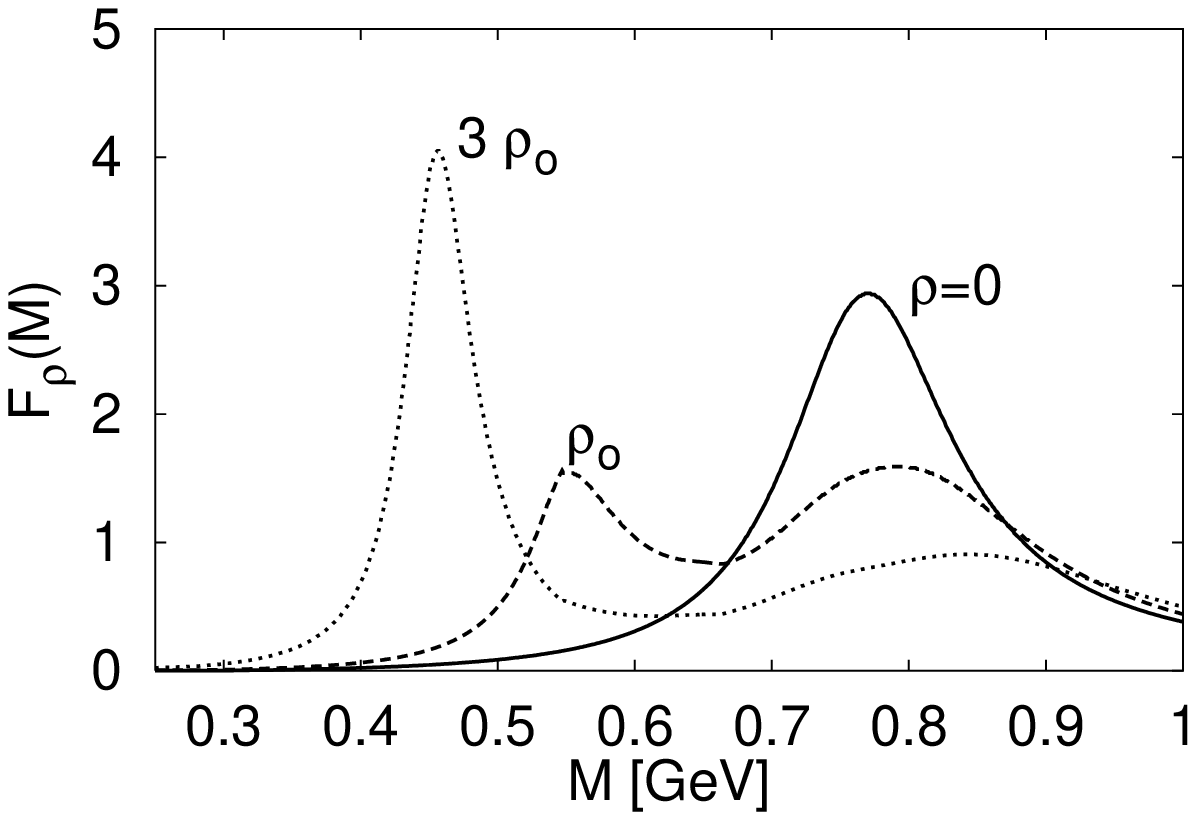,height=52mm}
\epsfig{file=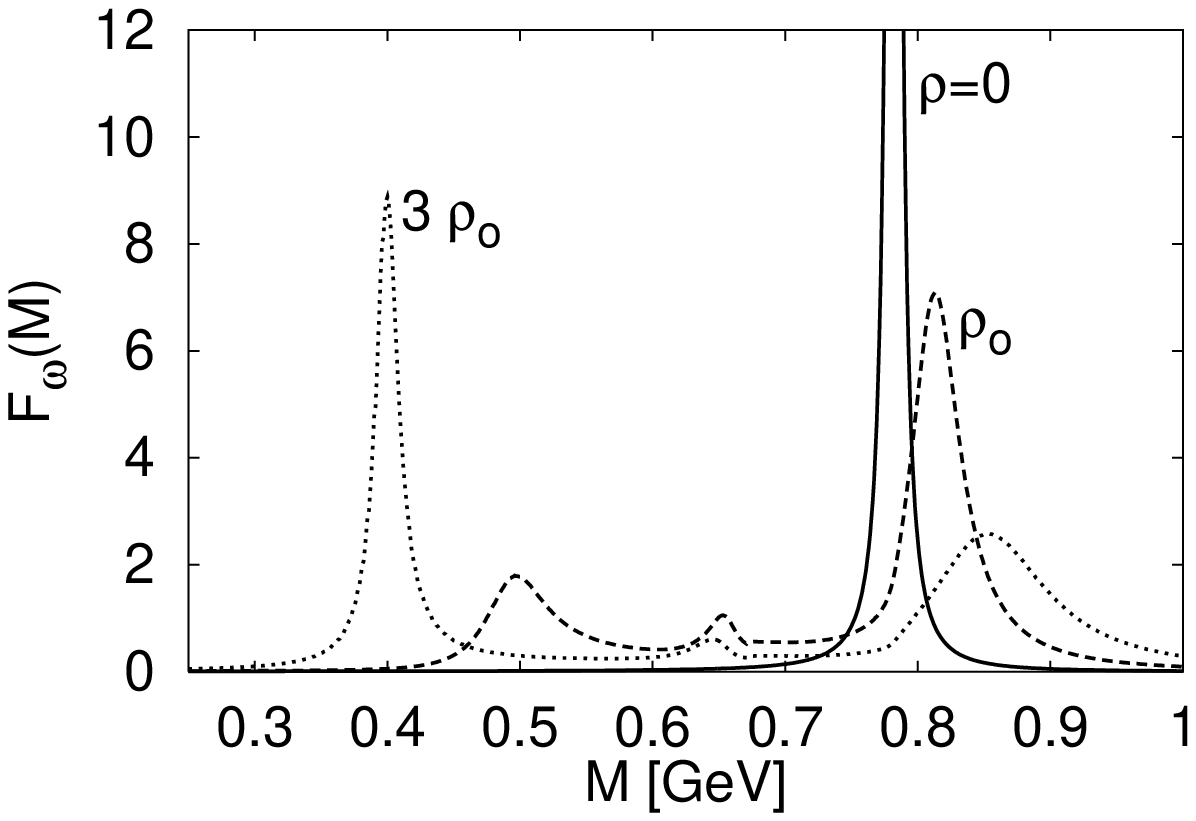,height=52mm}
\caption[]{Spectral functions of $\rho$ and $\omega$ mesons in nuclear
matter at densities $\rho = \rho_0$ and $3 \rho_0$ compared to those in
vacuum ($\rho = 0$).}
\label{spectfunc}
\end{center}
\end{figure}

We begin our discussion of the impact of in-medium modifications of
vector mesons on the nuclear equation of state and neutron star
properties with fig.\ \ref{spectfunc} which shows the spectral
functions of $\rho$ and $\omega$ mesons for three selected sample
densities.
These results are complementary to the recent calculations of ref.\
\cite{LWFnew}. 
Taking into account the photon induced reactions via a generalized
vector meson dominance 
the latter paper finds weaker medium effects for the $\rho$-meson
than established here. 
In the next step we
apply the two schemes introduced in section \ref{sec:schemes}, i.e.\
Brown-Rho scaling and modified spectral function method, to determine
the
\begin{figure}[tb]
\begin{center}
\leavevmode
\epsfig{file=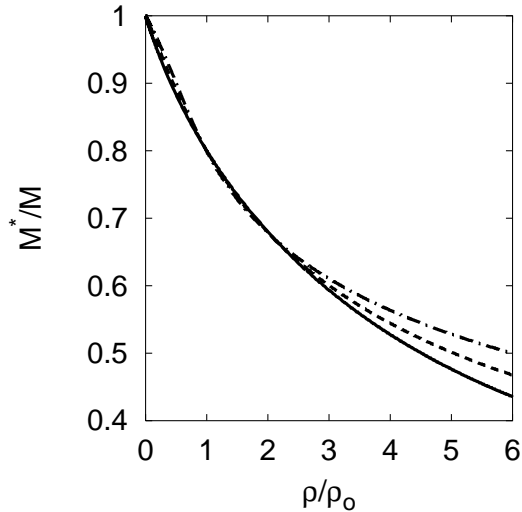,width=70mm}
\caption{Effective nucleon mass in isospin-symmetric nuclear matter
determined for three scenarios: no in-medium modifications of vector
mesons (solid line), medium modifications of vector mesons
computer for the spectral function method (dashed) and Brown-Rho
scaling scheme (dotted line).}
\label{mstar}
\end{center}
\end{figure}
medium induced modifications of vector mesons and calculate the
effective nucleon mass in matter, $M^*$, as well as the nuclear
equation of state over a considerable range of densities. The outcome
is shown in figs.\ \ref{mstar} and \ref{EOSnm}, respectively.  It is
evident from fig.\ \ref{mstar} that in-medium effects do not alter the
overall density dependence of $M^*$ in dense hadronic matter: the
effective nucleon mass drops monotonically with increasing density, as
known from standard calculations \cite{weber99:book}. At the density
of ordinary nuclear matter, $\rho_0$, $M^*$ is basically unchanged
from the value where in-medium effects on the vector mesons are
ignored.  Deviations however begin to show up at densities that are
three times $\rho_0$ or higher and grow monotonically with
density. This feature, together with the pronounced stiffening of the
equation of state caused by the in-medium effects, shown in fig.\
\ref{EOSnm}, renders such modifications very important for neutron
star properties. We will come back to this issue in greater detail in
section \ref{sec:nss}.  The stiffening of the equation of state is a
consequence of the fact that both energy density and pressure depend
on the vector meson masses as $\propto m^{-2}$. This proportionality
characterizes the high-density regime of the equation of state, which
is dominated by the exchange of vector mesons. Quantitatively, the
in-medium vector-meson masses $< m^{-2}_V>$ were computed
self-consistently, with the weights entering in the expectation values
given by the spectral functions.
\begin{figure}[tb]
\begin{center}
\leavevmode
\epsfig{file=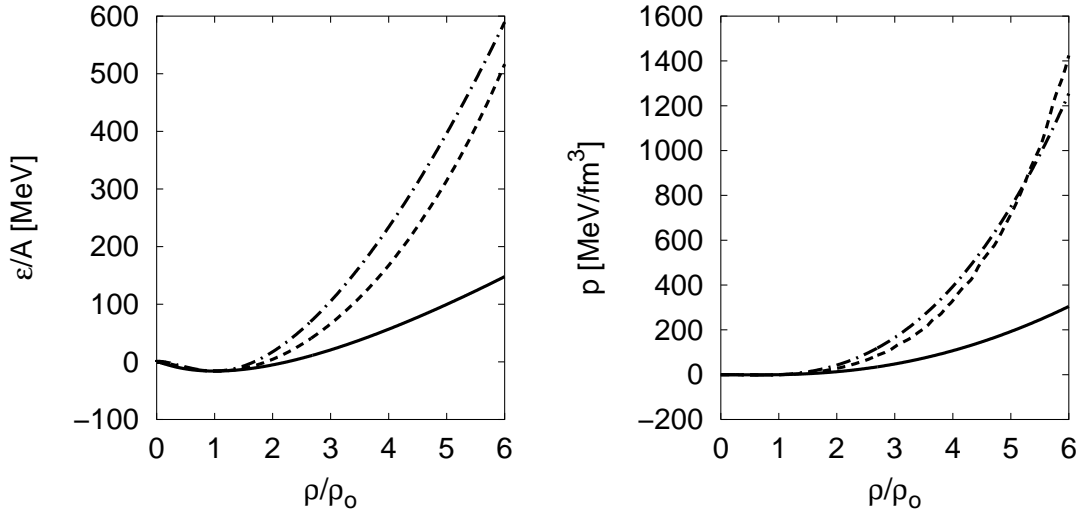,width=150mm}
\caption{Equation of state of isospin-symmetric nuclear matter.
The labeling of the curves is the same as in fig.\ \ref{mstar}}
\label{EOSnm}
\end{center}
\end{figure}

\section{Neutron star structure}\label{sec:nss}

In this section we explore the implications of in-medium modifications
of vector mesons for the structure of neutron stars.  The properties
of such objects provide important constraints on the behavior of dense
matter, for any model for the nuclear equation of state which fails to
accommodate neutron star masses of at least the mass of the
Hulse-Taylor radio pulsar PSR~1913+16, $(1.444\pm 0.003) \msun$, and
the rotational periods of the two fastest presently known neutron
stars, 1937+21 and 1957+20, each one rotating at 1.6~ms (620~Hz), can
definitely be ruled out as a viable candidate for the true equation of
state \cite{weber99:book}.

From model calculations, it is known that neutron stars possess rather
complex interior structures \cite{weber99:book}.  Only in the most
primitive conception, a neutron star is constituted from neutrons.  At
a more accurate representation, a neutron star contains neutrons ($n$)
and a certain number of protons ($p$) in chemical equilibrium,
\begin{equation}
 n \longleftrightarrow p + e^- + \overline{\nu} \, ,
\end{equation}
which implies for the corresponding chemical potentials that
\begin{equation}
\mu_n = \mu_p + \mu_e + \mu_{\bar\nu} \, .
\label{eq:chem}
\end{equation}
Since neutrinos do not accumulate in neutron stars older than a few
seconds, we set $\mu_{\bar\nu}=0$.  The electric charges of the
protons in the star must be balanced to very high precision by
electrons ($e^-$), which leads to the additional constraint for the
densities of protons and electrons 
\begin{equation}
\rho_p = \rho_e  \, .
\label{eq:cn}
\end{equation}
If the particles in the center of a neutron star would arrange
themselves in a way other than dictated by eq.\ (\ref{eq:cn}), gravity
would not be able to bind a neutron star.

\begin{figure}[tb]
\begin{center}
\leavevmode
\epsfig{file=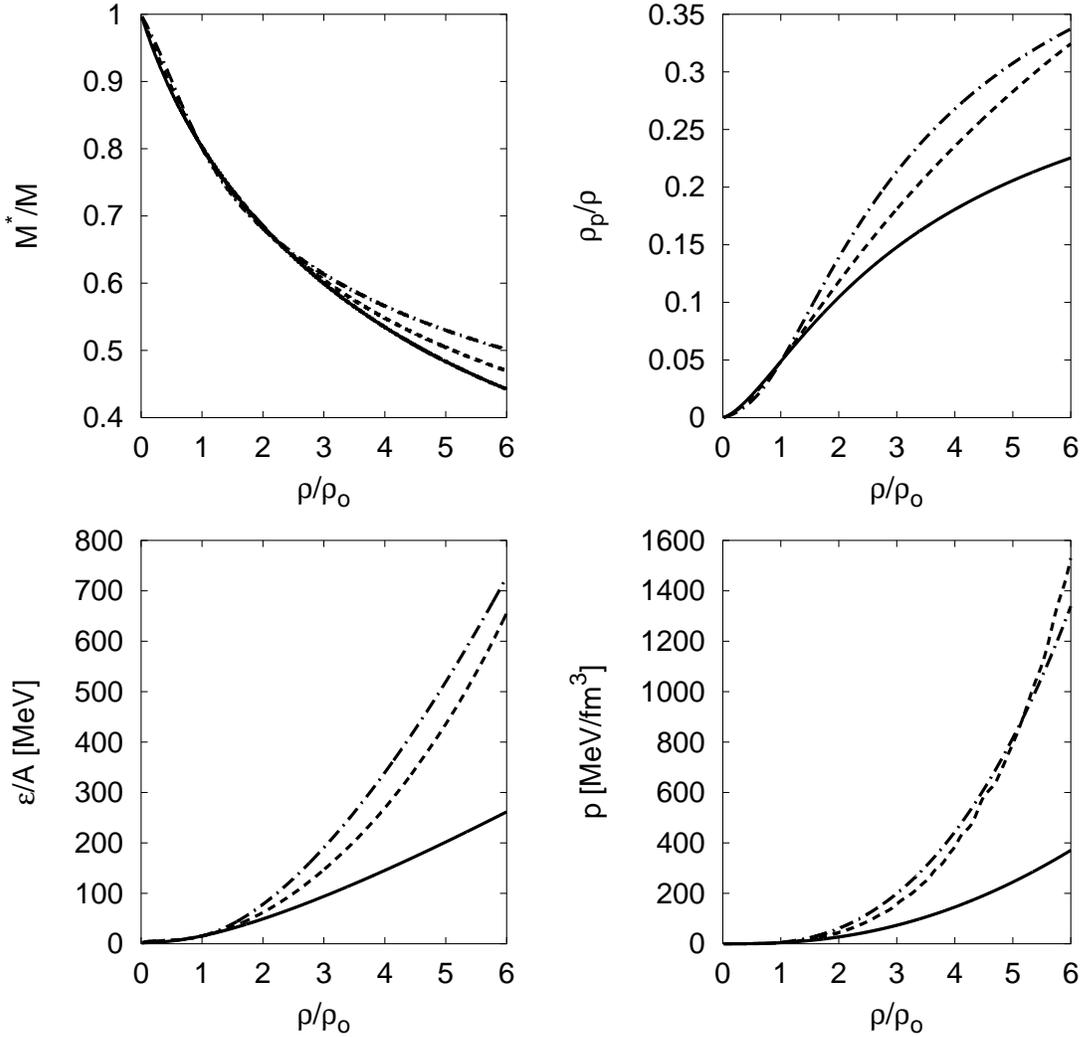,width=150mm}
\caption{Equation of state of chemically equilibrated neutron star
matter.  The labeling of the curves is the same as in fig.\
\ref{mstar}}
\label{EOSbe}
\end{center}
\end{figure}

Figure \ref{EOSbe} shows the equation of state of neutron star matter
computed from the lagrangian of eq.\ (\ref{eq:lag}) subject to the two
neutron star matter constraints (\ref{eq:chem}) and (\ref{eq:cn}).  As
for the nuclear matter case shown in fig.\ \ref{EOSnm}, the
free-particle case clearly provides the least pressure of all three
models. Yet the equation of state is stiff enough to accommodate
neutron stars as heavy as about $2\, \msun$, as will be discussed in
more detail below. Another issue concerns the relative proton fraction
in neutron star matter, shown in the upper-right panel of fig.\
\ref{EOSbe}, which is of central importance for the cooling of neutron
stars. As pointed out in \cite{lattimer91:a}, if $\rho_p/\rho$ becomes
larger than the critical fraction of about 11\%, the so-called direct Urca
process,
\begin{equation}
n \longrightarrow p + e + \nu_e \, ,
\label{eq:durca}
\end{equation}
where neutrons can transform to protons and electrons without the need
of a bystander particle, becomes possible in neutron stars. Otherwise
neutron stars were to cool via the less efficient standard Urca process,
\begin{equation}
n + n \longrightarrow n + p + e + \nu_e \, .
\label{eq:surca}
\end{equation}
The direct Urca process is crucial for the temperature evolution of
neutron stars as it speeds up the cooling of such objects considerably
in comparison with the standard Urca process. Our models for the
equation of state predict that the process in (\ref{eq:durca}) can
take place at densities around $2.3 \, \rho_0$ if in-medium effects on
vector mesons are ignored. Accounting for them, on the other hand,
brings this threshold down to densities as low as $1.5 \, \rho_0$.
Because of the rather flat density profiles in the inner parts of
neutron stars, this implies that the cores in neutron stars which
exhibit the direct Urca process are several kilometers bigger for an
equation of state accounting for in-medium effects.  As an example of
a neutron star that may require very fast cooling as driven by the
direct Urca process (\ref{eq:durca}) we quote pulsar PSR~1929+10
\cite{weber99:book}.

\begin{figure}[tb]
\begin{center}
\leavevmode
\epsfig{file=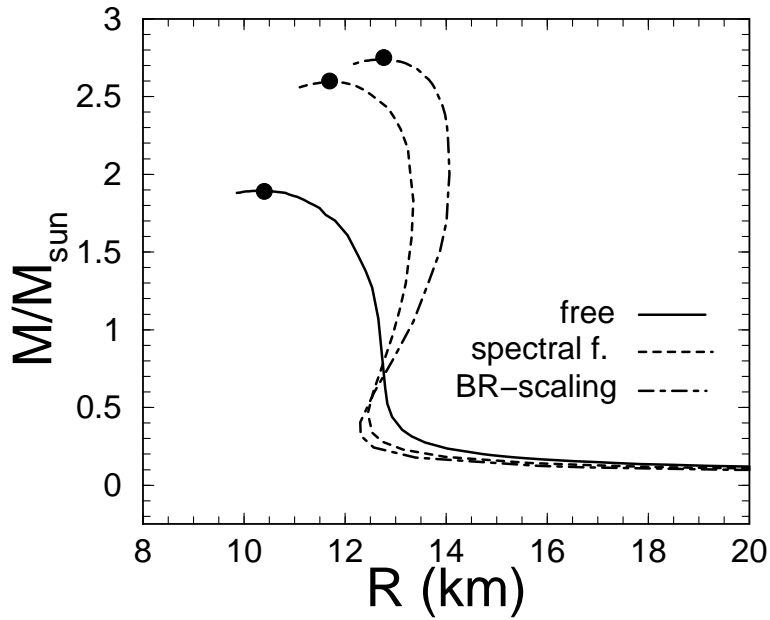,width=100mm}
\caption{Non-rotating neutron star mass versus radius, computed for
the EOSs of this paper. The solid dots denote the heaviest neutron
star of each sequence.}
\label{starradius}
\end{center}
\end{figure}

The stiffness/softness of the equation of state above nuclear matter
density intimately manifests itself in the interplay between radius,
mass, and central density of a neutron star, as discussed next.  We
begin with the mass-radius relationship of non-rotating neutron stars
computed by solving the Tolman-Oppenheimer-Volkoff equations, which
describe the properties of spherically symmetric stars that are in
general relativistic, hydrostatic equilibrium \cite{weber99:book}. The
outcome is shown in fig.\ \ref{starradius}, which reveals that our
collection of equation of states predicts limiting gravitational
neutron star masses between about $2\, \msun$ and $3\, \msun$, with
the high-mass end obtained for the equations of state that
include in-medium effects.  This range accommodates even the heaviest
neutron stars that may exist in stellar binary systems.  Examples of
which are the X-ray pulsar Vela X-1, whose mass is
$M=1.87^{+0.23}_{-0.17} \, \msun$ \cite{kerkwijk00:a}, and the burster
Cygnus X-2 whose mass is $M=(1.8\pm 0.4) \, \msun$ \cite{orosz00:a}.
Indications for the possible existence of very heavy neutron stars,
with masses around $2\,\msun$, may also come from the observation of
quasi-periodic oscillations in luminosity in low-mass X-ray binaries
\cite{klis00:a}. These mass values are larger than the typical
$(1.36\pm 0.08)\, \msun$ masses found in neutron star binaries,
presumably due to accreted matter. It is evident that neutron stars
that heavy, provided the mass determinations turn out to be robust,
would require extremely stiff equations of state. The inclusion of
in-medium effects of vector mesons would alter the equation of state
in the right direction.

\begin{figure}[htb]
\begin{center}
\epsfig{file=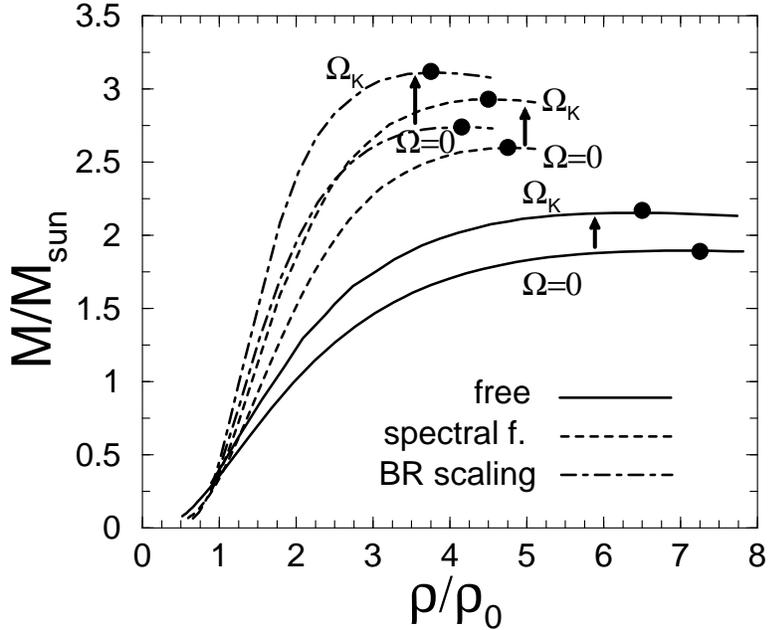,width=100mm}
\leavevmode
\caption{Neutron star mass versus central density, computed for the
EOSs of this paper.  The arrows indicated the rotation-induced mass
increase due to rotation at the Kepler frequency, $\okgr$. The solid
dots denote the heaviest star of each sequence.}
\label{stardensity}
\end{center}
\end{figure}

Another very important conclusion follows from the general feature
that stiff equations of state make neutron stars less dense
\cite{weber99:book}. This effect can be quite dramatic, depending of
the stiffness of the equation of state.  As shown in fig.\
\ref{stardensity}, the models computed in this paper predict central
densities on neutron stars of canonical mass of around $3\, \rho_0$ if
no in-medium effects are taken into account. This value drops down to
$\sim 1.5$ to $2\, \rho_0$ if in-medium effects are taken into
account. Therefore, if the medium effects as computed in this paper
have their correspondence in the full treatment of dense hadronic
matter, then the consequences for the composition of neutron stars of
masses $M \sim 1.5 \, \msun$ would be quite dramatic, for it leaves
basically no room for the existence of novel degrees of freedom, such
as quark matter or boson condensates, in the centers of such
objects. Only neutron stars with masses close to their limiting masses
(solid dots in fig.\ \ref{stardensity}) may possibly hide these novel
phases of matter.

Rapid neutron star rotation constitutes another important constraint
on the stiffness/softness of the nuclear equation of state.  The two
most rapidly rotating neutron stars presently known, PSR~1937+21 and
PSR~1957+20, rotate at 1.58~ms, which corresponds to about 620
rotations per second. Such high rotation rates constitute a
considerable fraction of the Kepler frequency, $\okgr$, at which mass
shedding from the star's equator sets it. Evidently, mass shedding
sets an absolute limit on rapid rotation which cannot be overcome by
any stably rotating object. In oder to determine $\okgr$ for a given
model for the equation of state one has to go beyond the
Tolman-Oppenheimer-Volkoff treatment and solve Einstein's field
equations
\begin{equation}
 {R}^{\mu\nu} - {1\over 2} g^{\mu\nu} R = 8 \pi
 T^{\mu\nu}(\rho,P(\rho)) \; ,
\label{eq:einstein}
\end{equation}
self-consistently in combination with the general relativistic
expression describing the onset of mass-shedding at the star's equator
\cite{weber99:book},
\begin{equation}
 \okgr = \omega + \frac{\omega^\prime}{2\psi^\prime} + e^{\nu -\psi}
  \sqrt{ \frac{\nu^\prime}{\psi^\prime} + \Bigl(\frac{\omega^\prime}{2
  \psi^\prime}e^{\psi-\nu}\Bigr)^2 } \;  .
\label{eq:okgr}
\end{equation}
The quantities $R^{\mu\nu}$, $g^{\mu\nu}$, and $R$ denote the Ricci
tensor, metric tensor, and Ricci scalar (scalar curvature),
respectively. The dependence of the energy-momentum tensor
$T^{\mu\nu}$ on the equation of state is indicated in eq.\
(\ref{eq:einstein}).  The quantities $\omega$, $\nu$, and $\psi$ in
eq.\ (\ref{eq:okgr}) denote the frame dragging frequency of the local
inertial frames, and the time and space-like metric functions,
respectively.  The primes denote derivatives with respect to
Schwarzschild radial coordinate, and all functions on the right are
evaluated at the star's equator.  All the quantities on the right hand
side of eq.\ (\ref{eq:okgr}) depend also on $\okgr$, so that it is not
an equation for $\okgr$, but a transcendental relationship which the
solution of the equations of stellar structure, resulting from eq.\
(\ref{eq:einstein}), must satisfy if the star is rotating at its
Kepler frequency.

\begin{figure}[htb]
\begin{center}
\leavevmode
\epsfig{file=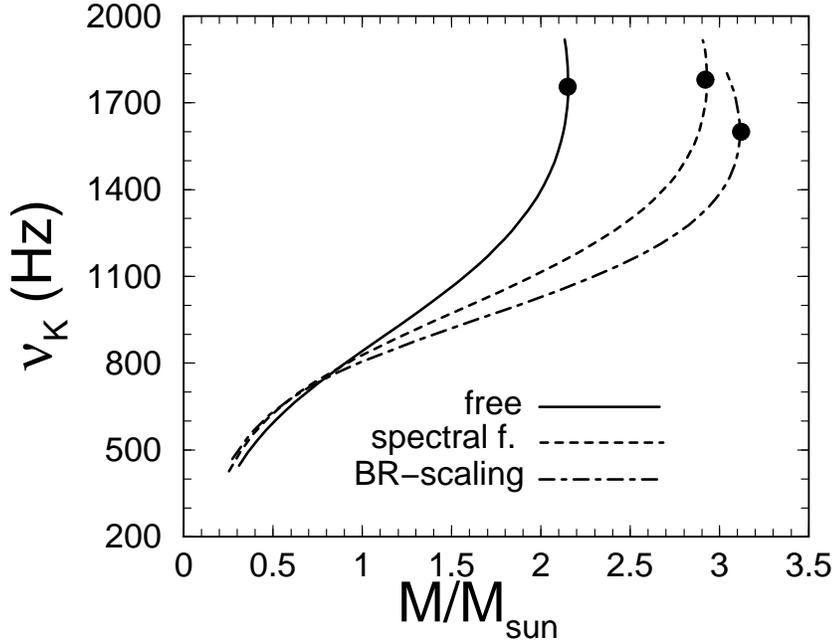,width=100mm}
\caption{Kepler frequency versus rotational neutron star mass,
computed for the EOSs of this paper.  The solid dots denote the
heaviest neutron star of each sequence.}
\label{starfreq}
\end{center}
\end{figure}
The numerical outcome for $\okgr$ is shown in fig.\ \ref{starfreq},
where we plot $\nuk$ ($\equiv 2 \pi / \okgr$) as a function of
rotational stellar mass, $M$.  For neutron stars of canonical mass, $M
\sim 1.4\, \msun$, one reads off from this figure that such stars can
perform up to $\sim 1000$ rotations per second before breaking up into
pieces. This comfortably accommodates the spin frequencies of the two
fastest rotating neutron stars presently known. Rapid rotation in
combination with stellar masses provides a double constraint on the
equation of state, since the successful model for the nuclear equation
of state must account for the smallest observed rotational neutron
star periods as well as the heaviest neutron star masses.  The
constraints on the equation of state become the more stringent the
smaller a neutron star's rotational period and the larger its mass.
Searches for such neutron stars are being performed at several radio
observatories. It is evident from fig.\ \ref{starfreq} that only very
compact neutron stars, compressed to small radius values, will be able
to withstand rapid rotation rates that are significantly higher than
600~Hz. Neutron stars at or close to the limiting mass of a given
stellar sequence are such candidates. Depending on stiffness/softness
of the equation of state, these may rotate at frequencies as large as
1500 to 1700~Hz, which corresponds to rotational periods between 0.7
and 0.4 milliseconds, respectively.  The only other class of stars
understood to withstand such rapid rotation are the strange stars,
which should replace neutron stars if 3-flavor strange quark matter is
the true ground-state of the strong interaction
\cite{madsen88:a,madsen97:bsky}, which, despite of many years of
research, is still an open issue. As a final point, we point out that
rapid rotation stabilizes a neutron star against gravitational
collapse. Figure \ref{stardensity} reveals that the mass that can be
supported due to rapid rotation may be up to $\sim 20$\% higher than
non-rotating mass value.  It is striking that the rotation-induced
mass increase is accompanied by a reduction in central star density,
contrary to what one would expect from simple mass loading onto a
star. The explanation is provided by the centrifugal pressure that
builds up inside a rotating star. For rapidly spinning neutron stars,
this contribution is quite significant so that considerably less
interior pressure (typically 30 to 40\%, as can read off from fig.\
\ref{stardensity} in combination with fig.\ \ref{EOSbe}) must be
provided by the equation of state itself. As already mentioned in
connection with the discussion of the properties of non-rotating
neutron stars, lower densities at the center of a neutron star rival
with exotic particle processes predicted for high-density matter, like
the formation of boson condensation and the transition into quark
matter.

\section{Conclusions}

This paper presents an investigation of the effect of in-medium
modification of vector mesons on the nuclear equation of state and the
properties of neutron stars. The latter serve to test the
compatibility of in-medium effects with observed data.  We find that
in-medium modifications reduce effectively the masses of vector mesons
which in turn stiffens the equation of state considerably.  This has
several intruigung implications for the structure and composition of
neutron stars. Firstly, because of the stiffening of the equation of
state, even the most massive neutron stars, like Vela X-1
($M=1.87^{+0.23}_{-0.17} \, \msun$) and the burst source Cygnus X-2
($M=(1.8\pm 0.4) \, \msun$), can be easily supported by equations of
state accounting for medium-modified vector meson
properties. Secondly, knowledge of the limiting neutron star mass is
key in order to identify black hole candidates. That is, if the mass
of a compact companion of an optical star is determined to exceed the
limiting mass of a neutron star, it must be a black hole. The limiting
mass of stable neutron stars in our theory is between 2.6 and $2.8\,
\msun$.  Hence any binary heavier than these values would be low-mass
black holes. The third striking point is that the computed neutron
star models have rather low central densities, just a few times the
density of ordinary nuclei for canonical neutron star masses around $M
\sim 1.5 \, \msun$. If this should withstand more elaborate future
treatments, it would leave basically no room for the existence of
novel degrees of freedom such as quark matter or boson condensates in
the centers of neutron stars. Such objects would then merely consist
of chemically equilibrated neutrons and protons.

\ack
This work were partly done while one of the author (Gy. Wolf) stayed
in JAERI as a STA fellow, and supported by Hungarian Research Foundation
(OTKA) grants T 26543 and T 30855 grant.


\section*{References}

\begin{thebibliography}{99}
\bibitem{CERES} CERES, G.~Agakichiev {\em et al.},
\Journal{\PRL}{75}{1272}{1995}; \Journal{\PLB}{422}{405}{1998};
\Journal{\NPA}{661}{23c}{1999}
\bibitem{Helios3} HELIOS-3, M. Masera {\em et al.},
\Journal{\NPA}{590}{93c}{1992}
\bibitem{likobrown} G.Q.~Li, C.M.~Ko and G.E.~Brown,
\Journal{\PRL}{75}{4007}{1995}
\bibitem{Brown-Rho} G.E.~Brown and M.~Rho, \Journal{\PRL}{66}{2720}{1991}
\bibitem{RCW} R.~Rapp, G.~Chanfray, J.~Wambach,
\Journal{\NPA}{617}{472}{1997}
\bibitem{weber99:book} F.~Weber, {\it Pulsars as Astrophysical Laboratories
for Nuclear and Particle Physics}, High Energy Physics, Cosmology and
Gravitation Series (IOP Publishing, Bristol, Great Britain, 1999)
\bibitem{Drees} A.~Drees, \Journal{\NPA}{610}{536c}{1996}
\bibitem{Lutz} M. Lutz, B. Friman, Ch. Appel, \Journal{\PLB}{474}{7}{2000}
\bibitem{Rapp} R.~Rapp, R. Machleidt, J.W. Durso, G.E. Brown,
\Journal{\PRL}{82}{1827}{1999}
\bibitem{LWF}  B. Friman, M.F.M. Lutz and Gy. Wolf, (Proceedings of the 
International Workshop on Gross Properties of Nuclei and Nuclear
Excitations, Hirschegg, Austria, 2000, ed. by H. Feldmeier $p.161$)
\bibitem{LDT} W.~Lenz, \Journal{\ZP}{56}{778}{1929};
C.D.~Dover, J.~H\"ufner and R.H.~Lemmer,
\Journal{\AP}{66}{248}{1971}; M.~Lutz, A.~Steiner and W.~Weise,
\Journal{\NPA}{574}{755}{1994}
\bibitem{LWFnew}  M.F.M. Lutz, Gy. Wolf, B. Friman, {\tt nucl-th/0112052} 
\bibitem{lattimer91:a} J.~M.~Lattimer, C.~J.~Pethick, M.~Prakash, and
P.~Haensel, \Journal{\PRL}{66}{2701}{1991}
\bibitem{kerkwijk00:a} M.~H.~ Van Kerkwijk, {\it Neutron Star Mass
Determinations}, to appear in Proc.\ ESO Workshop on Black Holes in
Binaries and Galactic Nuclei, Garching (Sept.\ 1999), eds.\ L.\ Kaper,
E.\ P.\ J.\ van den Heuvel, P.\ A.\ Woudt, Springer-Verlag
({\tt astro-ph/0001077})
\bibitem{orosz00:a} J.~A.~Orosz and E.~Kuulkers, Mon.\ Not.\ R.\ Astron.\
Soc.\ (in press)
\bibitem{klis00:a} M.~van der Klis, {\it Millisecond Oscillations
in X-Ray Binaries}, to appear in Ann.\ Rev.\ Astron.\ Astrophys.\
(2000)
\bibitem{madsen88:a} J.~Madsen, \Journal{\PRL}{61}{2909}{1988}
\bibitem{madsen97:bsky} J.~Madsen, AIP Conference Proc.\ 412,  Big Sky,
Montana, 1997, ed.\ by T.\ W.\ Donnelly (American Institute of
Physics, New York, 1997) p.\ 999
\end{thebibliography}
\end{document}